\documentstyle[preprint,aps]{revtex}

\begin{document}
%\draft

\title{Quasiparticle damping in two-dimensional
superconductors with unconventional pairing}
\author{M. L. Titov, A. G. Yashenkin and D. N. Aristov}
\address{Petersburg Nuclear Physics Institute, Gatchina, St.
Petersburg 188350, Russia.}
\date{\today}
\maketitle

\begin{abstract}
We calculate the damping of excitations due to
four-fermionic interaction in the case of two-dimensional
superconductor with nodes in the spectrum. At zero
temperature and low frequencies it reveals gapless
$\omega^3$ behavior at the nodal points. With the
frequency increasing the crossover to the normal-state
regimes appears. At high frequencies the damping
strongly depends on details of a normal-state
spectrum parametrization. Two important particular cases
such as the models of almost free and tight-binding
electrons are studied explicitly and the characteristic
scales are expressed through the model-free parameters of
the spectrum at the nodal points. The possibility of
crossover in temperature dependence of damping in the
superconducting phase is discussed.  \end{abstract}

%%%%%%%%%%%%%%%%%%%%%%%%%%%%%%%%%%%%%%%%%%%%%%%%%%%%%%%%%%%%

\pacs{}

\narrowtext
\section{Introduction} An assumption that the pairing state
in the high-temperature superconductors is the
$d$-wave-like one is widely discussed nowadays. \cite{1}
Although up to now it apparently cannot be treated as
the uniquely established fact there exist a large body of
experimental data which supports it. \cite{2}  Such a
pairing would result in presence of ``nodal'' points in
the superconducting excitations spectrum in whose
vicinities it has gapless phonon-like character. It
should essentially modify the damping behavior comparing
to the $s$-wave pairing state.

The quasiparticle damping owing to impurities has been
extensively studied in literature. \cite{3} However, this
is not the only channel, though, maybe, the more
interesting one. The excitations scattering by phonons and
also by each others always contribute to the damping, as
well.

In the present paper we evaluate the quasiparticle damping
in two-dimensional (2D) superconductors with $d$-wave-like
pairing appeared due to their four-fermionic interaction. It
may be the on-site Coulomb repulsion in the small-$U$
Hubbard model, or the residual part of the interaction
which provides the pairing in the strong correlations
scheme. \cite{4,5}

In the BCS case, the damping at zero temperature  has
the threshold at $\omega = 3 \Delta$. We show that for the
gap function with nodes it demonstrates gapless $\omega^3$
low-frequency behavior at the nodal points. The results
obtained for high frequencies depend on details of Fermi
surface (FS) parametrization. The damping for the circular
FS will increase quadratically while for the spectrum
with Van-Hove singularities the additional linear regime
appears. Being generalized onto finite temperatures, this
results allow us to conclude about existence of crossovers
in temperature dependence of damping in the superconducting
state.

The outline of this paper is as follows. In Sect. II we
obtain general expression for damping of superconducting
excitations due to their interaction. In Sect. III we
evaluate it for the case of 2D superconductor with nodes in
the spectrum. Since the high-frequency behavior of damping
depends strongly on FS parametrization, in the consequent
subsections we analyze three principal cases:  (i) the case
of circular FS and also the case of FS of varying curvature
when the gap node intersects it (ii) at the almost flat
piece or (iii) in the vicinity of saddle-point of the
spectrum. At last, in Sect. IV we generalize and discuss
the results obtained.

%%%%%%%%%%%%%%%%%%%%%%%%%%%%%%%%%%%%%%%%%%%%%%%%%%%%%%%%%%%%

\section{General formalism}

It is convenient to represent the Hamiltonian
${\cal H} = {\cal H}_o + {\cal H}_{i}$ in the Nambu
notation {\begin{eqnarray} {\cal H}_{o}&=&\sum_{{\bf
k}}\Psi_{{\bf k}}^{+} (\xi_{{\bf k}}\tau_{3}+\Delta_{{\bf
k}}\tau^{-}+\Delta_{{\bf k}} ^{*}\tau^{+})\Psi_{{\bf k}}, \\
{\cal H}_{i}&=&\frac{1}{2}\sum_{\bf {k,k',q}}V_{\bf
{k+q,k'-q;k,k'}} (\Psi_{\bf {k+q}}^{+}\tau_{3}\Psi_{\bf
k})(\Psi_{\bf {k'-q}}^{+}\tau_{3}\Psi_{\bf k'}). \nonumber
\end{eqnarray}} Here $V_{{\bf k_1}{\bf k_2}; {\bf k_3} {\bf
k_4}}$ is the interaction, $\Psi_{{\bf k}}^{\, +} = (
c_{{\bf k \uparrow} }^{\, \dagger} \,,\, c_{{\bf -k
\downarrow}^{}} )$ stands for the spinor with  $c_{{\bf k}
\sigma}^{\, \dagger}$ being the fermionic creation
operator, $\xi_{\bf k}$ and $\Delta_{\bf k}$ are the
normal-state conduction band and the gap function (GF),
respectively, and $\tau^{\pm} = \frac{1}{2} \left( \tau_1
\pm i \tau_2 \right)$, where $\tau_{i}$ denote the Pauli
matrices.

Let us define the bare matrix Green function as follows
\begin{equation}
{\hat G}^{(0)} = \left(
{\begin{array}{cc}
{G^{(0)}} & -{F^{(0)}} \\
{F^{(0)}}^{+} & {G^{(0)}}^{+}, \end{array}} \right)
\end{equation}
where normal $G$ and anomalous $F$ Green functions are
written as
\begin{eqnarray} G_{{{\bf
k}}}^{(0)} (i \omega_{n}) \, =- \, {\displaystyle \frac{i
\omega_{n} + \xi_{{{\bf k}}}}{\omega^{2}_{n} +
\varepsilon_{{{\bf k}}}^2}, } \qquad F_{{{\bf k}}}^{(0)} (i
\omega_{n})  \, = \, {\displaystyle \frac{\Delta_{{{\bf
k}}}} {\omega^{2}_{n} + \varepsilon_{{{\bf k}}}^2}. }
\end{eqnarray}
Here $i \omega_{n} = 2 \pi i (n + \frac{1}{2}) T$ is the
imaginary Matsubara frequency and the excitation spectrum
is given by $\varepsilon_{{\bf k}}=[\, \xi_{{\bf
k}}^2+|\Delta_{{\bf k}}|^2]^{1/2}$.

Within this notation the Dyson equation obtains the form
${\hat G}^{-1} = {\hat G}^{(0) \, -1} - \hat \Sigma$,
with the diagonal self-energy operator $\hat \Sigma = {\rm
diag} \, (\Sigma , \Sigma^{+})$.

The first nonvanishing contribution to $\Sigma$ is given by
\begin{eqnarray} \Sigma({\bf k},i\omega_{n})
&=& T^2\sum_{\bf 1,2,3} V_{1,2;3,0} V_{3,0;1,2} \, \delta
_{1,2;3,0}\nonumber \\ &\times &\left[ \, G^{(0)}_{\bf 1}
G^{(0)}_{\bf 2} G^{(0)}_{\bf 3} + F^{(0)}_{\bf 1}
{F^{(0)}_{\bf 2}}^{+} {G^{(0)}_{\bf 3}}^{+} \right],
\end{eqnarray} where the indices ${\bf 1,2,3}$ label the
intermediate frequencies and momenta and ${\bf 0}$
corresponds to the external one.

After the analytic continuation $i \omega_{n} \to \omega +
i \delta$ for the imaginary part of the self-energy at
$T=0$ and $\omega > 0$ we get
\begin{eqnarray} \FL {\rm
Im}\Sigma({\bf k},\omega) &=&-\pi\!\sum_{{\bf q} ,{\bf p}}
\int\limits_{0}^{\omega}\!\!dx\!\!\!\!
\int\limits_{0}^{\omega-x}\!\!\!\!dy\,{V_{\bf k,q+p-k;q,p}
V_{\bf q,p;k,q+p-k}} \nonumber \\ &\times &
\delta\,(x\!-\!\varepsilon_{{\bf q}})\>
\delta\,(y\!-\!\varepsilon_{{\bf p}})\>
\delta\,(x\!+\!y\!-\!\omega\!+\!  \varepsilon_{{\bf q}+{\bf
p}-{\bf k}}) \nonumber \\ &\times &[u^2_{{\bf
q}}u^2_{{\bf p} }v^2_{{\bf q} +{\bf p} -{\bf k}}-u_{{\bf q}
}u^{*}_{{\bf p} }v_{{\bf q}}v^{*}_{{\bf p} }u^2_{{\bf q}
+{\bf p} -{\bf k}}], \end{eqnarray}
where the coefficients
of Bogolyubov transform are defined by \begin{equation}
u^{2}_{{\bf q}} = \frac{1}{2} \left(1 + \frac{\xi_{ {\bf
q}}}{\varepsilon_{\bf q}} \right), \qquad v^{2}_{{\bf q}} =
\frac{1} {2} \left( 1 - \frac{\xi_{{\bf
q}}}{\varepsilon_{\bf q}} \right). \end{equation}

%%%%%%%%%%%%%%%%%%%%%%%%%%%%%%%%%%%%%%%%%%%%%%%%%%%%%%%%%%%%

\section{Calculations of damping}

In this section we will evaluate Eq. (5) in the case of 2D
superconductor with unconventional pairing assuming
different model parametrizations for the Fermi surface.

We shall suppose that the gap function has nodal lines
$|\Delta_{{\bf q}}|=0$ in the 2D Brillouin zone. Let these
lines intersect Fermi surface at a set of the points
${\bf q}={{\bf q}}_{n}$. Then in the vicinities of such
points one can write \begin{equation} \begin{tabular}{rcl c
rcl} $\xi_{\bf q} $&$ \simeq $&$ {\bf v}_F {\bf \bar q},
$&$ \qquad \qquad $&$ \Delta_{\bf q} $&$ \simeq $&$ {\bf
v}_g {\bf \bar q} ;$ \\ & & &$ \qquad \qquad $& & &  \\
${\bf v}_F $&$ = $&$ { \displaystyle \left. \frac{ \partial
\xi_{{\bf q}} }{ \partial {\bf {\bar q}} } \right|_{ {\bf
{\bar q}}=0 } }, $&$ \qquad \qquad $&$ {\bf v}_g $&$ = $&$
{ \displaystyle \left. \frac{ \partial \Delta_{{\bf q}} }{
\partial {\bf {\bar q}}} \right|_{{\bf {\bar q}}=0 } },$
\end{tabular} \end{equation} where ${\bf \bar q} = {{\bf
q}} - {{\bf q}}_{n}$ and the index $n$ labels the nodal
point. We see that the excitation spectrum near these
points has gapless linear character.

In general, the quasiparticle damping due to their
interaction has to reveal threshold feature in frequency in
the superconducting state. The threshold corresponds to the
minimal energy of the excitation needed to decay into three
quasiparticles in the intermediate state. For the gap
function with nodes however it disappears at the nodal
points of the spectrum. Consequently, the damping at this
points appear to be gapless. Taking it into account, we
restrict ourselves in treatment of damping in the vicinities
of nodal points only.

The straightforward calculations at $\omega \to 0$ yield
\begin{equation}
\begin{array}{rcl} & {\rm Im} \, \Sigma \, ({\bf
k}_{n},\omega) \simeq - \lambda_{n} \omega^3 & \nonumber \\
& &  \nonumber \\ & \lambda_{n} = {\displaystyle \sum_{l,k}
\frac{V^2_{n;lk} \,\, {\cal F}_{lk}}{|{\bf
v}^{l}_{F}\!\times \!  {\bf v}^{l}_{g}||{\bf
v}^{k}_{F}\!\times \!  {\bf v}^{k}_{g}|} } & \\ & &
\nonumber \\ & V^2_{n;lk}={\displaystyle \lim_{{\bf k},{\bf
p},{\bf q} \to {\bf k}_{n},{\bf p}_{l},{\bf q}_{k}} {V_{\bf
k,q+p-k;q,p}V_{\bf q,p;k,q+p-k}}}, & \nonumber \end{array}
\end{equation} where the indices $l,k$ run over all the
nodal points. Here ${\cal F}_{lk}$ are some dimensionless
positive factors. The diagonal terms of the matrix ${\hat
{\cal F}}$ (i.e., referring to a single nodal point) are
always numbers of order of unity while the off-diagonal
ones could have additional smallness for particular FS
parametrizations via their dependence on spectrum
parameters. The principal contribution to the damping at
low frequencies comes from excitations with momenta
lying in the small elliptic vicinities of nodal points
(see Fig. \ref{1}) wherein linear approximation for the
spectrum works well.

Note that we obtain the universal cubic frequency
dependence of the excitations damping at the nodal points
without appealing to any fixed parametrization for
spectrum. By the way, such dependence means the stability
of nodes with respect to this kind of interaction.

At the same time, some features of the model may, in
principle, lead to different power laws. In particular, the
damping would increase with frequency slower than
${\omega}^3$ if at least one of spectrum zeroes (i)
appeared due to tangency of FS and GF node, or (ii)
coincided with van-Hove singularity in the band
structure and/or (iii) with cusp- or discontinuity-point
of the nodal line of the gap function.  All these cases
are obviously quite exotic.

When the frequency increases, the integration in Eq. (5)
cannot be restricted to the nodal vicinities and such
characteristics as general shape of FS,
curvature of FS at the nodal points, existence of
van-Hove singularities in the spectrum, $etc.,$ should be
taken into account. One can distinguish three qualitatively
different cases: (i) the circular FS and also the FS of
varying curvature with the nodal points located (ii) far
from or (iii) at the sharp curvature maximum. We shall
illustrate the above cases by examples which would allow us
to examine in detail the intermediate asymptotics and to
express the corresponding crossover scales through the
parameters of the models used.

\subsection{Circular Fermi surface}

Putting the lattice parameter to be unity, we
write the circular FS as follows: $\xi_{{\bf q}} = q^2/2m -
\varepsilon_{F}$. For definiteness, we choose the $d_{x^2 -
y^2}$-wave gap function, ${\Delta}_{{\bf q}}=
{\Delta}/{2} \, [\, \cos q_x - \cos q_y \,]$. \cite{5,6}
In line with Eq. (8) we get:  \begin{eqnarray} {{\rm
Im}\Sigma({\bf k}_{n},\omega)} \simeq -{\left({V \over
{\varepsilon}_{F}}\right)}^2
\frac{\omega^3}{\Delta^2},\quad
\omega\ll\frac{\Delta^2}{\varepsilon_F}, \end{eqnarray}
where $V^2 \sim \max V^2_{n;ls}$. The principal
contribution to ${\rm Im}\Sigma$ is obtained when all the
momenta ${\bf p}$, ${\bf q}$ and ${{\bf k}}_{n}$ are almost
parallel and belong to the vicinities of the nodal points
(see Fig. \ref{1}).

At higher frequencies $\Delta^2 / \varepsilon_F \ll
\omega \ll \varepsilon_F$ we find
\begin{eqnarray} {{\rm
Im}\Sigma({\bf k}_{n},\omega)} \simeq -{\left({V \over
\varepsilon_{F}}\right)}^2 \frac{\omega^2}{\varepsilon_{F}}
\ln \left[ \min \left( \frac{ \omega \,
\varepsilon_{F}}{\Delta^2} \>
;\frac{\varepsilon_{F}}{\omega}\right) \right]
\end{eqnarray}
To obtain Eq. (10), the curvature of FS was taken into
account (see Appendix). The crossover from the cubic
dependence to the quadratic one appears at
$\omega \sim {\Delta^2}/{\varepsilon_{F}}$. It can
be qualitatively explained on the following way. One can
show that, when $\omega \ll {\Delta^2}/{\varepsilon_{F}}$,
the areas of integration defined by arguments of
$\delta$-functions in Eq. (5) are located near the nodal
points (see Fig. \ref{1}) where the spectrum can be well
approximated by ellipses. When $\omega \sim
{\Delta^2}/{\varepsilon_{F}}$, the areas of integration
propagate along the whole Fermi surface so that not only
the vicinities of the nodal points principally contribute
to ${\rm Im} \Sigma$. For ${\Delta^2}/{\varepsilon_{F}}
\ll \omega \ll \Delta $, we obtain the quadratic
intermediate asymptotic regime with unconventional
logarithmic factor of the form $\ln \,(\omega \,
\varepsilon_{F} / {\Delta}^2)$ which
stems from integration over the FS. At frequencies higher
then the gap value $\Delta \ll \omega \ll \varepsilon_{F}$
information about the superconducting state disappears, and
the conventional 2D Fermi-liquid result \cite{7} is
restored. Note that principal contributions to ${\rm
Im} \Sigma$ in the latter case again come from almost
parallel ${\bf k}$, ${\bf p}$ and ${\bf q}$.

\subsection{Tight-binding model, $d_{x^2 - y^2}$-wave
pairing}

\begin{figure}
\caption{The disposition of the momenta which give the
principal contribution to Im$\Sigma$ at $\omega \gg \mu$
in the case of nested FS. The crosses label the
saddle-points. The nodal points are located (a) far away
and (b) near the saddle ones.}
\label{2}
\end{figure}
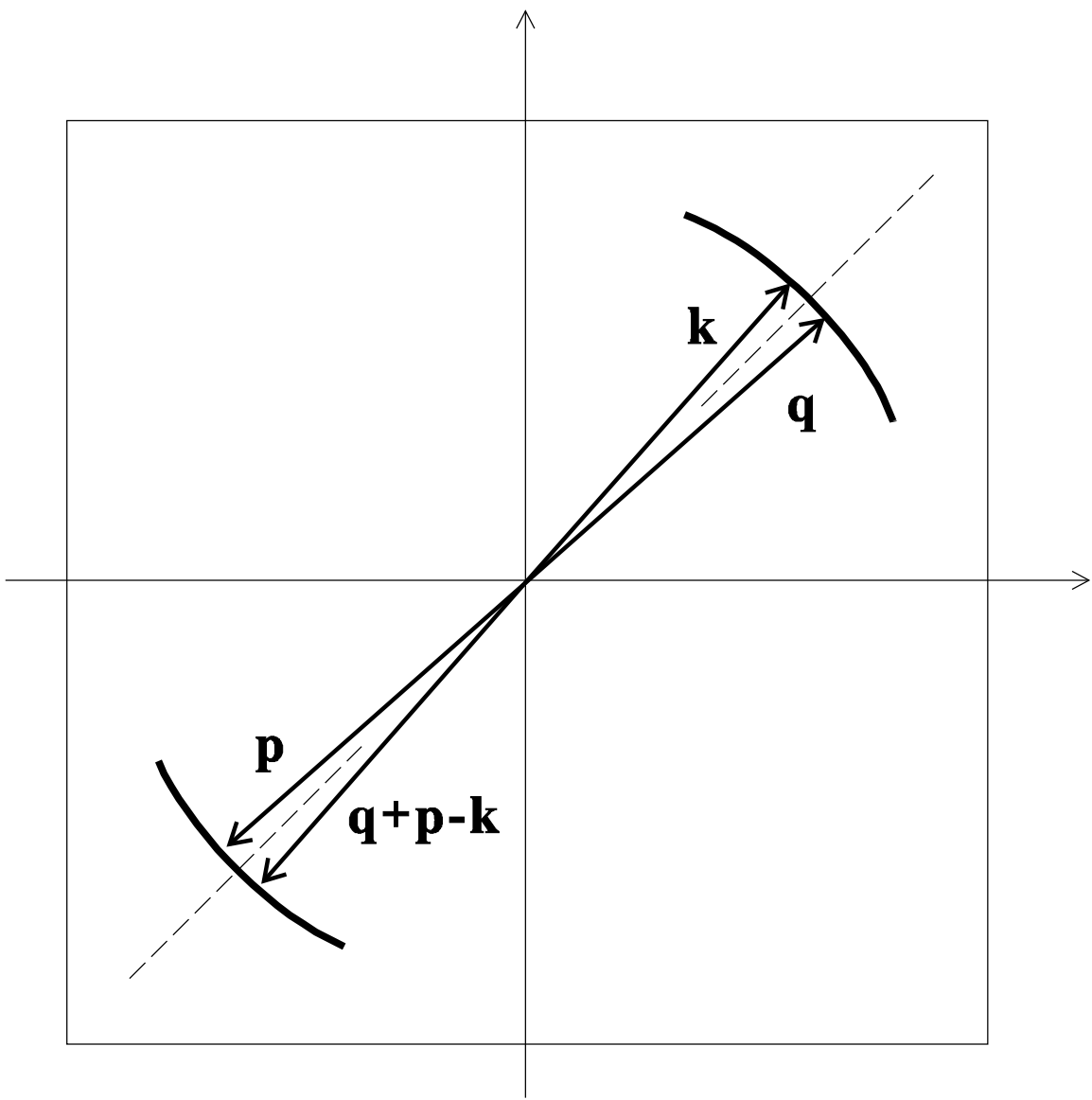
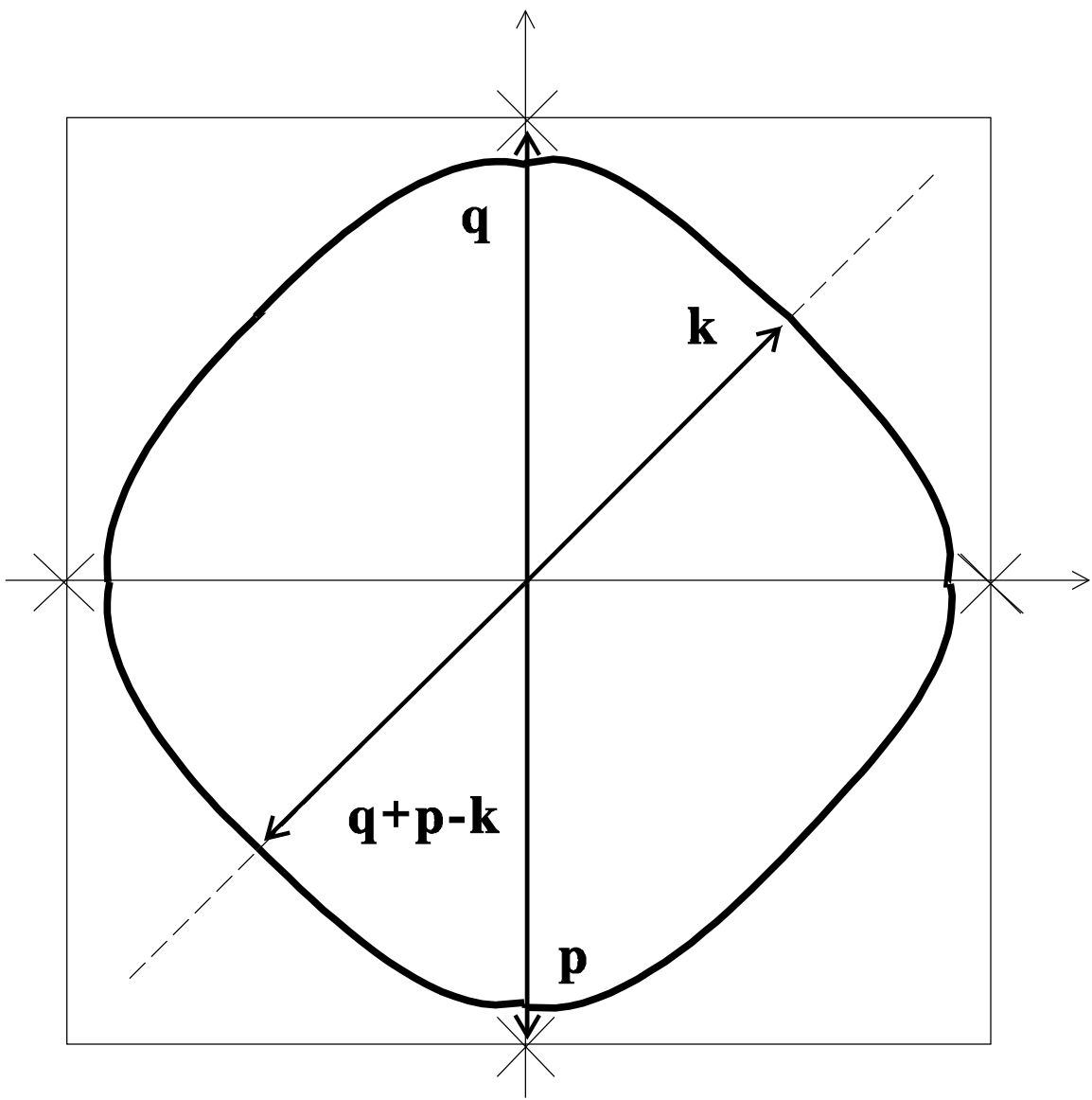
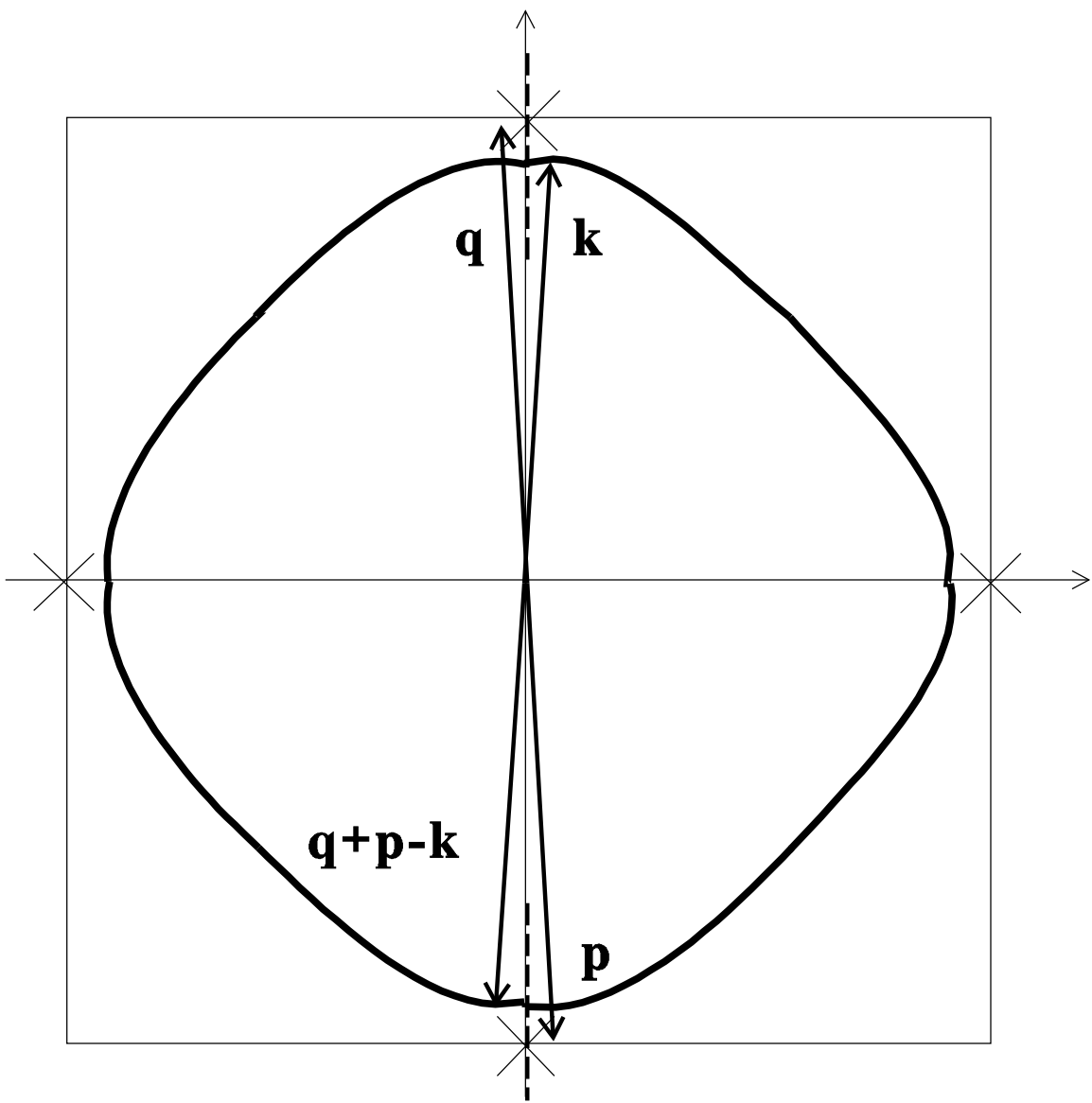

Now we will demonstrate how the deformation of FS affects
the damping. As before we shall consider $d_{x^2-y^2}$ gap
function. Let us adopt the simplest tight-banding form for
the normal-state dispersion:  $\xi_{{\bf q}}=-2t \, [\,\cos
q_{x} +\cos q_{y} \,]+\mu$.  Here $t$ is the
nearest-neighbors hopping integral and $\mu$ is the
chemical potential measured from the half-filling. Note
that for small enough $\mu \ll t$ Fermi surface has
long almost flat pieces near the points of intersection
with the nodal lines of the $d_{x^2-y^2}$ gap function (see
Fig. \ref{2}a).

Under the above conditions one can easily rewrite the
low-frequency asymptotics of damping by substituting
${\varepsilon}_{F} \to t$ in the parenthesis and
${\varepsilon}_{F} \to \mu$ in other parts of Eq. (9). The
range of validity for such expression is restricted by the
condition $\omega \ll \min \left({\Delta^2}/{\mu},
\Delta \right)$. Next, in order to obtain expression
for damping at $\min \left({\Delta^2}/{\mu}, \Delta
\right) \ll \omega \ll \max \left( \mu , \Delta \right)$,
the same substitutions in Eq. (10) should be done.
Obviously, all the statements concerning the areas of
principle contributions still hold. Finally, at $\max
\left( \mu , \Delta \right) \ll \omega \ll t$ we obtain
\begin{equation}
{{\rm Im}\Sigma({\bf k}_{n},\omega)} \simeq
-{\left( \frac{V}{t}\right)}^2 \, \omega
\label{11}
\end{equation}

Thus, the additional linear-in-$\omega$ high-frequency
regime appears in this case. It occurs since at $\omega
\gg \max \left( \mu , \Delta \right)$ the areas of
integration in Eq. (5) engage the saddle-points of the
normal-state spectrum located at $(\pm \pi , 0)$ and $(0 ,
\pm \pi)$. For the normal state crossover to the linear
behavior has been demonstrated in \cite{8}.

It follows from the aforementioned that there is no room
for linear regime of damping at $\mu \sim t$ while the
crossover from the cubic dependence to the quadratic one
occurs at $\omega \sim {\Delta}^2 / \mu$. This case
can be treated as the one of slightly deformed circular
FS without additional characteristic scales. When $\mu$
decreases (i.e., close to the half-filling) the long almost
flat pieces of the FS appear as well as the saddle-points
in the normal-state spectrum come into play.  For $\mu \ll
\Delta \ll t$ there co-exist three ranges with cubic,
quadratic and linear frequency dependencies matching at
$\omega \sim {\Delta^2}/{\mu}$ and $\omega \sim \mu$,
respectively. At last, for very small $\mu \ll \Delta$
linear approximation for spectrum works well unless
the ellipses touch the saddle-points. Consequently, the
range of quadratic dependence in the latter case disappears
and there exist the only crossover scale $\omega \sim
\Delta$.

\subsection{Tight-binding model, $d_{xy}$-wave pairing}

Finally, let us consider the situation when the nodal
points lie at the points of maximum of FS curvature. To
illustrate this case we write the normal-state spectrum in
the tight-binding form and choose the gap function as
$d$-wave with $xy$-symmetry, \cite{9} ${\Delta}_{{\bf q}} =
\Delta \, \sin q_{x} \, \sin q_{y}$ (see Fig. \ref{2}b).
Then the straightforward calculations at $\omega \ll \mu$
yield the formulas similar to those obtained in previous
subsection:  \FL\begin{eqnarray} {{\rm Im}\Sigma({\bf
k}_{n},\omega)} \simeq -{\left({V \over \mu}\right)}^2
\frac{\omega^3}{\Delta^2}, \qquad
\omega\ll\frac{\Delta^2\mu}{t^2}.
\end{eqnarray}
When $\Delta^2 \mu /{t^2} \ll \omega \ll \mu$ we have
\FL\begin{equation}
{{\rm Im}\Sigma({\bf k}_{n},\omega)} \simeq -{\left({V
\over t}\right)}^2 \frac{\omega^2}{\mu}\ln \left[ \min
\left( \frac{\omega \, t^2}{\Delta^2\mu}\>
;\frac{\mu}{\omega} \right) \right].
\end{equation}
Again, we obtain the crossover from the cubic regime to the
quadratic one. We see that close to the half-filling (at $
\mu \ll t$) the energy scale of this crossover (${\Delta^2
\mu}/{t^2}$) is much less than that previously considered.
It is connected with large curvature of FS near the nodal
points so that the linear approximation for the spectrum
breaks down much earlier. Note that this narrowing of the
$\omega^3$ region is accompanied by the strong parametric
enhancement of Im$\Sigma$ by factor ${t^2}/{\mu^2}$.

At higher frequencies $\mu \ll \omega \ll t$ we get
\FL\begin{equation}
{{\rm Im}\Sigma({\bf k}_{n},\omega)} \simeq
-{\left({V \over t}\right)}^2 \> \omega \>
\ln \left[ \min \left( \frac{\omega}{\mu} \> ;
\frac{t}{\omega} \right) \right]
\label{14}
\end{equation}
Comparing this formula with formula found in previous
subsection we see that the additional logarithmic
factor exists stemming from closeness of saddle-point
to the nodal one.

Note also that under $\mu \to 0$ we deal with
poor-defined excitations since as it follows from our
calculations ${\rm Im} \Sigma \sim \omega \ln \omega$ for
all $\omega$ in this case.  From Eq. (14) and
Kramers-Kronig relations one can find that the
quasiparticles are well-defined if \begin{equation} \left[
\frac{V}{t}\ln\frac{t}{\mu} \right]^2 \leq 1 \end{equation}
which is nothing but the criterion of absence of spin
density wave. We note here that the same criterion arises
in the case considered in Subsect. B, as well. The
additional logarithmic enhancement stems in that case from
the self-energy behavior near the ``corners'' of FS (cf.
\cite{8}).

%%%%%%%%%%%%%%%%%%%%%%%%%%%%%%%%%%%%%%%%%%%%%%%%%%%%%%%%%%%%

\section{Discussion and Conclusions}

All the results presented in previous section may be
rewritten in an universal form. To do that more compactly
let us introduce the following variables
$a={v_{g}^2}/({v_{F}{\cal K}})$ and $b=\min
\left(v_{g},{v_{g}}/{{\cal K}}\right)$, where ${\cal
K}$ is the curvature of FS at the nodal point. We define
also $\alpha$ as the angle between ${\bf v}_{F}$ and ${\bf
v}_{g}$. Then we may write down the following formula:

\widetext
\begin{equation}
{{\rm Im}\Sigma({\bf k}_{n},\omega)}
\simeq -\frac{V^2}{v_{F}^2v_{g}^2\sin^2\alpha}
\times \left\{
\begin{array}{lc}
\omega^3 \, , &\omega \ll \min \, (a,b);  \\
\!\!\omega^{2}a \ln \left[ \min
\left({\displaystyle \frac{\omega}{a} ,\frac{b^2}{a\omega}}
\right) \right], & \min \, (a,b) \ll \omega\ll \max
\left(b,{\displaystyle\frac{b^2}{a}} \right); \\
\omega b^2\times\left( 1\ \mbox{or} \ \ln {\displaystyle
\frac{a\omega}{b^2}} \right)\, , &
{\displaystyle\frac{b^2}{a}\ll\omega},
\end{array} \right.  \end{equation}
\narrowtext
\noindent
This result is
valid until $v_{g}\ll v_{F}$ and $\alpha \neq 0$, as it
takes place in all principal cases. The additional
extrafactor $\ln \, ({a \omega}/{b^2})$ at high frequencies
$\omega \gg {b^2}/{a}$ appears only when the nodal point
lies near the saddle one. It is easy to see that all
previous results are reproduced by the following
substitutions (i) ${\cal K}
\sim 1$, $v_{g} \sim \Delta$, $v_{F} \sim \varepsilon_{F}$
(assuming $q_F \sim 1$) for circular FS, (ii) ${\cal K}
\sim \mu /t$, $v_{g} \sim \Delta$, $v_{F} \sim t$ and (iii)
${\cal K} \sim \sqrt{t/\mu}$, $v_{g} \sim
\Delta\sqrt{\mu/t}$, $v_{F} \sim \sqrt{\mu t}$ for FS
of varying curvature. We mention also that the quadratic
regime disappears completely when ${\cal K}\ll
{v_{g}}/{v_{F}}$ and that the intermediate crossover scales
differ principally as ${\cal K}<1$ and ${\cal K}>1$.

Note that we have evaluated the damping using spectrum
expansion near specific points up to the second order
terms in deviations from this points. This is obviously
valid for Fermi surfaces smooth enough.

We undertake our consideration for circular and simplest
tight-binding FSs. Meanwhile, in the adequate theory of
high-temperature superconductivity the
next-nearest-neighbor hopping terms in the
tight-binding scheme should be taken into account.
\cite{10} It is clear from our analysis how the picture
will be modified in this case. Say, for $d_{x^2 -
y^2}$-wave pairing the range of cubic regime will be
shrunken and a set of intermediate crossovers may appear in
the range of the quadratic one. However, they will differ
only by the logarithmic factor. Next, for the special
choice of the parameters $2t^{\prime} \approx t$ and $\mu
\simeq 4 t^{\prime}$ the saddle-points near $(0, \pm \pi)$
and $(\pm \pi, 0)$ appear to be the ``extended'' ones
(quasi-1D). \cite{11} Here $t^{\prime}$ is the
next-to-nearest neighbor hopping integral and $\mu$
controls the filling. This case should be treated
separately.

The above consideration was made for zero temperature. It
can be trivially generalized onto low finite temperatures
(less than $\omega$ and minimal frequency scale in (16)).
In the opposite limiting case (but not very close to
$T_{c}$) Eqs. 8-14 can be rewritten by substitution $T$ for
$\omega$. In particular, it leads to the cubic
$T$-dependence of damping for sufficiently small
temperatures: $\omega\ll T\ll\min \, (a,b)$ (the
temperature dependence of the gap function yields only the
corrections for this law).  In order to apply our results
to the normal state one should put simply $v_{g}=0$. Note
that divergence in the low-temperature asymptotics never
exist, since always $T_{c} \gg a$. If the crossover scales
$\min \, (a,b)$ or $\max\left(b,{b^2}/{a}\right)$ are found
below $T_{c}$, $T^2$ or $T$ regimes should appear in the
superconducting phase. When ${b^2}/{a} \gg T_{c}$, the
crossover from the conventional 2D FL behavior to the
linear one occurs in the normal state.

Our treatment has demonstrated, in particular, that the
``purely nodal'' regime for damping takes place in the
quite narrow frequency (temperature) range. It is
restricted from above by the much less scale than the gap
value as one could expect. It seems to be not a feature
of damping only, so that the same characteristic scale
should appear for other quantities such as density of
states, nuclear relaxation rate, penetration depth,
inelastic scattered neutron intensity near incommensurate
and internodal vectors, $etc$. Since, on the other hand,
the impurities scattering restricts this range from below,
one can estimate it as follows:
\begin{eqnarray} \Gamma_{imp} &\lesssim & (\omega ,T) \,
\lesssim \, \varepsilon_{n} \,
\left[\frac{v_{g}}{v_{F}}\right]^2,\nonumber \\
\varepsilon_{n}&=&\left(\frac{q_{F}^{2}}{2m}\right)_{{\bf
 q}={\bf q_{n}}}, \end{eqnarray}
where $\Gamma_{imp}$ is the damping due to impurities.
The estimates show that the range defined by Eq. (17)
appear to be quite narrow for the high-$T_c$ compounds.
It may even disappear if $\Gamma_{imp}$ will be large
enough ($\sim$10$meV$).

In conclusion, we consider the damping of excitations
in the 2D superconductor with nodal points in the spectrum.
We show that at $T=0$ this damping demonstrates gapless
cubic-in-$\omega$ low-frequency behavior at the nodal
points. The results obtained for high frequencies depend on
details of Fermi surface parametrization. We have analyzed
two important particular FS parametrizations such as the
models of almost free and tight-binding electrons. We find
the frequency variation of damping and the crossover scales
for these cases. For smooth enough FS they may be expressed
through the model-free characteristics of the spectrum at
the nodal points. We argue that the excitations are
well-defined in the wide frequency range. We extend
our treatment to finite temperatures and study the
possibility of crossovers in temperature dependence of
damping below the superconducting transition.

\acknowledgments We thank S. V. Maleyev for useful
discussions and critical reading of the manuscript.
The financial support from the International Science
Foundation (Grant No. R3Y000) is acknowledged.

%%%%%%%%%%%%%%%%%%%%%%%%%%%%%%%%%%%%%%%%%%%%%%%%%%%%%%%%%%%%

\appendix
\section{}

Here we study Eq. (5) for  circular FS:  $\xi_{{\bf
q}}=q^2/2m-\varepsilon_{F}$. We shall assume that $V_{\bf
k,q;p,l}$ is finite in the important areas of the integrand
in Eq. (5). Also, let us take for a moment the
combination of $u-v$ coefficients in Eq. (5) to be
of order of unity. Then we should analyze the following
expression \begin{eqnarray} {\rm Im} \Sigma \simeq
V^2\!\!\int\!\!\!\int_{D} d^{2}{\bf q}\, d^{2}{\bf p}\>
\delta \, (\varepsilon_{{\bf q}}+\varepsilon_{{\bf
p}}-\omega+\varepsilon_{{\bf q}+{\bf p}-{\bf k}}),
\end{eqnarray} where D is defined by the condition
\begin{equation} \varepsilon_{{\bf q}}+\varepsilon_{{\bf
p}} \leq \omega. \end{equation} It is convenient to make the
substitution ${\hat q}=q^2-q_{F}^2$ , ${\hat
p}=p^2-q_{F}^2$ in Eq. (A1), then $d^2{\bf q}=d\varphi
d{\hat q}$, $d^2{\bf p}=d\psi d{\hat p}$, where $\varphi$
and $\psi$ are the corresponding polar angles.
Taking $q_{F} \sim 1$, we may represent the normal-state
dispersion as $\xi_{{\bf q}} \simeq \varepsilon_{F}{\hat
q}$. Besides, we expand the gap functions up to the linear
in angle terms $\Delta_{{\bf q}} \simeq \Delta\varphi$,
$\Delta_{{\bf p}} \simeq \Delta\psi$. Using such
parametrization and leaving the principal terms only, we
reduce Eq. (A1) to the following one \begin{eqnarray} \FL
{\rm Im} \Sigma &\simeq& V^2 \hspace*{-0.80cm}
\int\limits_{\ 0}^{\quad \min(1,\omega /\Delta)}
\hspace*{-0.90cm} \int d\varphi \, d\psi \hspace*{-0.30cm}
\int\limits_{\ 0}^{\quad \omega /\varepsilon_{F}}
\hspace*{-0.50cm} \int dq \, dp \; \delta \,\biggl(
\sqrt{q^2\varepsilon_{F}^{2}+\Delta^2 \varphi^2} \nonumber
\\ &&+\sqrt{p^2\varepsilon_{F}^{2}+\Delta^2\psi^2}
-\omega+\varepsilon_{F}\varphi\psi \biggr), \end{eqnarray}
where the limits of integration stem from inequality (A2).
This condition means, in particular, that at small $\omega$
momenta integrations in Eq. (A3) are restricted by the
elliptic areas whose centers lie at the nodal points. It is
convenient to pass therefore to the elliptic coordinates in
Eq. (A3): \begin{equation} \begin{tabular}{rcl c rcl} $r$ &
$=$ & $\sqrt{q^2\varepsilon_{F}^{2}+\Delta^2 \varphi^2}$ &
$\qquad$ & $R$ & $=$ &
$\sqrt{p^2\varepsilon_{F}^{2}+\Delta^2\psi^2}$ \\
\vspace*{0.20cm} $\alpha$ & $=$ & $\arccos
q\varepsilon_{F}/r$ & $\qquad$ & $\beta$ & $=$ & $\arccos
p\varepsilon_{F}/R$ \\ \end{tabular} \end{equation} Then it
transforms to \begin{eqnarray} \FL {\rm Im} \Sigma &\simeq&
\frac{V^2\omega^3} {\varepsilon_{F}^2{\Delta}^2} \quad
\hspace*{-0.80cm} \int\limits_{\ 0}^{\quad \min(1,\Delta
/\omega)} \hspace*{-0.90cm} \int  d\alpha \, d\beta \;
\hspace*{-0.20cm} \int\limits_{\ 0}^{\quad 1}
\hspace*{-0.30cm} \int dr \, dR \nonumber \\ && \times
\delta
\left(\frac{r+R-1}{rR}+\frac{\varepsilon_{F}}{\Delta^2}
\omega\alpha\beta \right)
\end{eqnarray}
This expression allows us to analyze all principal limiting
cases. \\ \noindent(1) $\omega\ll
{\Delta^2}/{\varepsilon_{F}}$. In this case we can
neglect the second term in $\delta$-function.  The integral
remained can be easily evaluated and the result is given by
Eq. (9) \\ \noindent(2)
${\Delta^2}/{e_{F}}\ll\omega\ll\Delta$. Here it
is convenient to rewrite $\delta$-function in Eq. (A5) as
follows \begin{equation}
\frac{\Delta^2}{\omega\varepsilon_{F}}\ \delta\,\left(
\alpha\beta + O\left(\frac{\Delta^2}{\omega\varepsilon_{F}}
\right) \right).  \end{equation} One can see that the
integration over one of the angles gives the logarithmic
divergence which is cut on the scale
${\Delta^2}/{\omega\varepsilon_{F}}$.  \\
\noindent(3)
$\Delta\ll\omega\ll\varepsilon_{F}$. Then we should
integrate the $\delta$-function in Eq. (A5) with respect to
the angles between the limits $0$ and
$\Delta / \omega$. After the trivial substitution
$(\alpha, \beta)$ $\mapsto ({\Delta}/{\omega})\,(\alpha,
\beta)$, we have \begin{equation}
\frac{\omega}{\varepsilon_{F}}\delta\left( \alpha
\beta+O\left(\frac{\omega}{\varepsilon_{F}}\right) \right)
\end{equation} Again, we have the logarithmic divergence,
which is cut now on the scale ${\omega}/{\varepsilon_{F}}$.
The last two cases we may be combined as Eq. (10).

It can be checked out that taking the combination of $u-v$
coefficients in Eq. (5) into account does not change our
estimates.

Other kinds of Fermi surface and gap functions considered
in the text can be analyzed on the similar manner.

%%%%%%%%%%%%%%%%%%%%%%%%%%%%%%%%%%%%%%%%%%%%%%%%%%%%%%%%%%%%

\begin{figure}
\caption{The disposition of the momenta which give the
principal contribution to Im$\Sigma$ at $\omega \to 0$.
The solid lines denote the Fermi surface and the dashed
lines correspond to the nodes of the gap function.}
\label{1}
\end{figure}
\end{document}